\documentclass{iau}
\usepackage{graphicx}
\usepackage{hyperref} 

\title[Magnetic fields of Sun-like stars] 
{Magnetic fields of Sun-like stars}

\author[Rim Fares]   
{Rim Fares
}

\affiliation{School of Physics and Astronomy, University of St Andrews, KY16 9SS, Scotland, UK\\email: {\tt rim.fares@st-andrews.ac.uk}}

\pubyear{2013}
\volume{302}  
\jname{Magnetic Fields Throughout Stellar Evolution}
\editors{P. Petit, eds.}
\begin{document}

\maketitle

\begin{abstract}
Magnetic fields play an important role at all stages of stellar evolution. In Sun-like stars, they are generated in the outer convective layers. Studying the large-scale magnetic fields of these stars enlightens our understanding of the field properties and gives us observational constraints for the field generation models. In this review, I summarise the current observational picture of the large-scale magnetic fields of Sun-like stars, in particular solar-twins and planet-host stars. I discuss the observations of large-scale magnetic cycles, and compare these cycles to the solar cycle.
\keywords{stars: magnetic fields -- stars: activity -- stars: individual
-- techniques: spectropolarimetry}
\end{abstract}

\firstsection 
\section{Introduction}
\label{sec:intro}

Magnetic fields are present at different scales in the universe, from planets to stars, galaxies and galaxy clusters. In the case of stars, they play an important role at all stages of stellar evolution, from the collapse of the molecular cloud, through the pre-main sequence and main sequence phases to more evolved stages including supernovae, white dwarfs and neutron stars. They influence and control a number of physical processes, such as accretion, diffusion, mass-loss, angular momentum loss, and turbulence. Thus, studying the characteristics and generation of stellar magnetic fields is a necessary step to increase our understanding of stellar evolution (and also planetary formation and evolution). 

The Sun is the closest - and thus the best studied - star. The discovery of its magnetic field goes back to the beginning of the 20th century when Hale, using the newly discovered Zeeman effect (\cite[Zeeman 1897]{1897ApJ.....5..332Z}), found that sunspots are magnetic features (\cite[Hale 1908]{1908ApJ....28..315H}). We now know that many observed features are due to magnetic fields, such as spots, faculae, and coronal mass ejection. 

The solar magnetic field evolves in time. Sunspots emerge in mid-latitude activity belts, and the latitudes of these activity belts migrate towards the equator on a timescale of 11 years. This gives the well know butterfly diagram. I will refer to this cycle as the activity cycle. However, the large-scale magnetic field of the Sun varies on a different timescale. The polarity of the field flips every 11 years, meaning that the large-scale cycle is actually 22 years. 

Activity cycles are observed on a number of other stars by studying activity proxies for magnetic fields such as CaII H\&K and X-rays (for reviews, see \cite[Baliunas et al. 1995]{1995ApJ...438..269B}; \cite[Baliunas et al. 1997]{1997ApJ...474L.119B};  \cite[Metcalfe et al. 2010]{2010ApJ...723L.213M}; \cite[Sanz-Forcada et 
al. 2013]{2013A&A...553L...6S}; \cite[Berdyugina 2005]{2005LRSP....2....8B}). Sun-like stars with outer convective layers like the Sun generate their magnetic field by dynamo mechanisms active in these outer layers (e.g. \cite{2011ApJ...731...69B}, \cite[Charbonneau(2010)]{2010LRSP....7....3C}  for a review on solar dynamo models). The study of the magnetic field of Sun-like stars therefore brings new insights and constraints to the current dynamo theories. This enhances our knowledge of the large-scale magnetic fields of stars, which allows us to test how `normal' the Sun is in a sample of Sun-like stars. The results I review here are for Sun-like stars of spectral types F, G and K, having masses between $\sim0.7$ and 1.5 M$_{\odot}$. They have different depths of the outer convective envelope. Comparing stars with different properties can lead to a better understanding of the dynamo generation of the field. 

\section{Magnetic Mapping}

In order to study large-scale stellar magnetic fields, one can examine the polarisation in the spectral lines. If a magnetic field is present where those lines are formed, due to the Zeeman effect, spectral lines will be polarised (the polarisation level depends on the magnetic sensitivity of the particular line). The polarisation properties depend on the position of the observer relative to the orientation of the magnetic field. For example, circular polarisation is sensitive to the line-of-sight component of the magnetic field (see \cite[Landi Degl'Innocenti \& Landolfi 2004]{2004ASSL..307.....L}). 

The aim of magnetic mapping is to reconstruct stellar large-scale magnetic field orientation, geometry and strength. When the star rotates, the observer sees different parts of the stellar disc. If those parts have different magnetic field distributions, the polarisation in the spectral lines will not be the same in spectra taken at different rotational phases. Thus, the technique used to reconstruct the large-scale magnetic field is a tomographic technique, like the one used in Magnetic Resonence Imaging. It is called Zeeman-Doppler Imaging (ZDI), and consists of inverting series' of circular polarised spectra into a magnetic topology, i.e. the distribution of magnetic fluxes and field orientations (\cite[Semel 1989]{1989A&A...225..456S}). Since the inversion problem is ill-posed, regularisation techniques are used to get a unique map. These techniches include maximum entropy (\cite[Brown et al. 1991]{1991A&A...250..463B}; \cite[Hussain et al. 2000]{2000MNRAS.318..961H}) and Tikhonov regularisation (\cite[Piskunov \& Kochukhov 2002]{2002A&A...381..736P}). The results presented here are mostly obtained using the maximum entropy method. The magnetic field is described by its radial poloidal, non-radial poloidal and toroidal components, all described using spherical harmonics expansions (\cite[Donati et al. 2006a]{2006MNRAS.370..629D}). ZDI is a powerful technique in recovering the large-scale magnetic field of the star, as well as its differential rotation. However, it has its limitations, because the small-scale fields are not resolved up to a certain limit, their signatures cancel out in some field geometries, and the field in dark spots is suppressed (see, e.g. \cite[Johnstone et al. 2010]{2010MNRAS.404..101J}).

Collecting polarised spectra is possible using spectropolarimeters, such ESPaDOnS on CFHT, its twin instrument NARVAL on TBL, and HARPSpol on the 3.6-m in La Silla (\cite[Donati et al. 2006b]{2006ASPC..358..362D}; \cite[Piskunov et al. 2011]{2011Msngr.143....7P}; \cite[Snik et al. 2011]{2011ASPC..437..237S}). The polarisation signature is extremely small ($\sim10^{-4}$). In order to increase the S/N ratio of the data, a multi-line technique called Least-Square Deconvolution (LSD) is used. It produces a mean profile with a higher S/N ratio than in single lines, depending on the number of lines used to calculate the mean profile (\cite[Donati et al. 1997]{1997MNRAS.291..658D}; \cite[Kochukhov et al. 2010]{2010A&A...524A...5K}).

\section{The Magnetic Topologies of Sun-like Stars}

In this section, I will present the results of different studies targeting Sun-like stars. Some of these stars were observed by the Bcool project\footnote{\url{http://bcool.ast.obs-mip.fr}}, a project aiming at studying the magnetic fields of Sun-like stars and solar twins. Another campaign targeted hot-Jupiter hosting stars. It aimed to investigate interactions between the planet and the star, as well as studying how the stellar field influences the environment in which the planet evolves. 

\begin{figure}
\centering
	\includegraphics[height=\textwidth,angle=90]{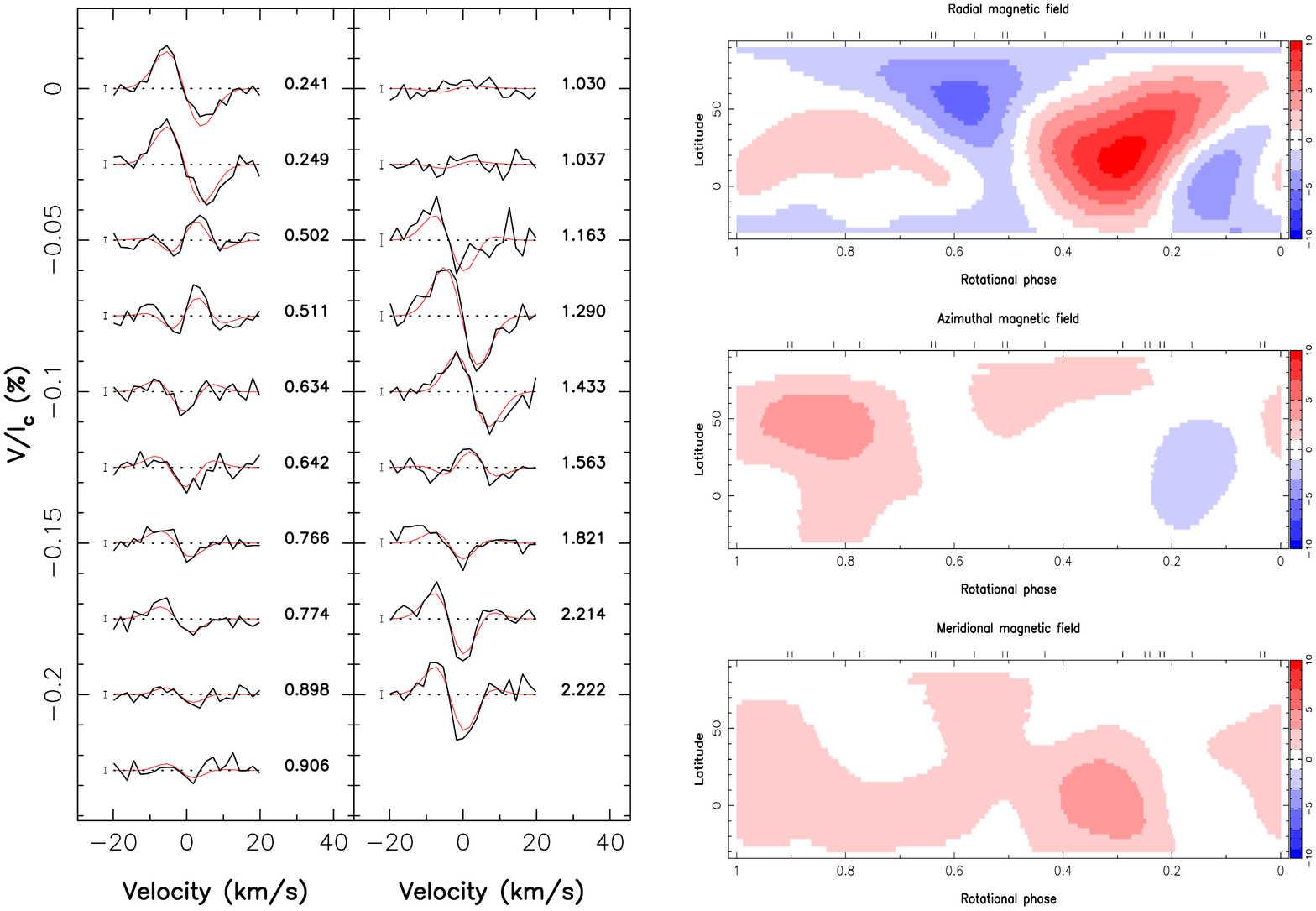}
\caption[]{Left panel: Circular polarization profiles of HD~179949 for 2009~September. The observed and synthetic profiles are shown in black and red respectively. On the left of each profile we show a $\pm1~\sigma$ error bar, while on the right, the rotational cycles are indicated. Right panel: The three components of the field in spherical coordinates are presented. Adapted from \cite{2012MNRAS.423.1006F}}
\label{fig:hd179949}
\end{figure}

HD~179949 is a F star with $\rm T_{eff} = 6120$~K (\cite[Nordstr{\"o}m et 
al. 2004]{2004A&A...418..989N}), $\rm M \sim 1.18$ M$_{\odot}$, $\rm v sini = 7.0\pm0.5$~km s$^{-1}$ (\cite[Valenti \& Fischer 2005]{2005yCat..21590141V}), $\rm P_{rot} = 7.6 - 10.3$ days, $d\Omega = 0.22$ rad day$^{-1}$ (\cite[Fares et al. 2012]{2012MNRAS.423.1006F}), and hosts a hot-Jupiter. The stellar activity was reported to be modulated by the planetary orbital period instead of the stellar rotation period during some epochs (\cite[Shkolnik et al. 2003]{2003ApJ...597.1092S};\cite[Shkolnik et al. 2005]{2005AJ....130..799S}; \cite[Shkolnik et al. 2008]{2008ApJ...676..628S}). This was interpreted as a possible stellar activity enhancement by the planet due star-planet interactions. \cite[Fares et al.(2012)]{2012MNRAS.423.1006F} observed this star during two epochs. Figure \ref{fig:hd179949} shows the circular polarisation LSD profiles they obtained in September 2009 (left panel) and the reconstructed map (right panel).  The three components of the magnetic field in spherical coordinates are shown. The mean magnetic field is 4 G, with 90\% of the energy in the poloidal component (mainly radial). 

\begin{figure}
\centering
	\includegraphics[height=\textwidth,angle=90]{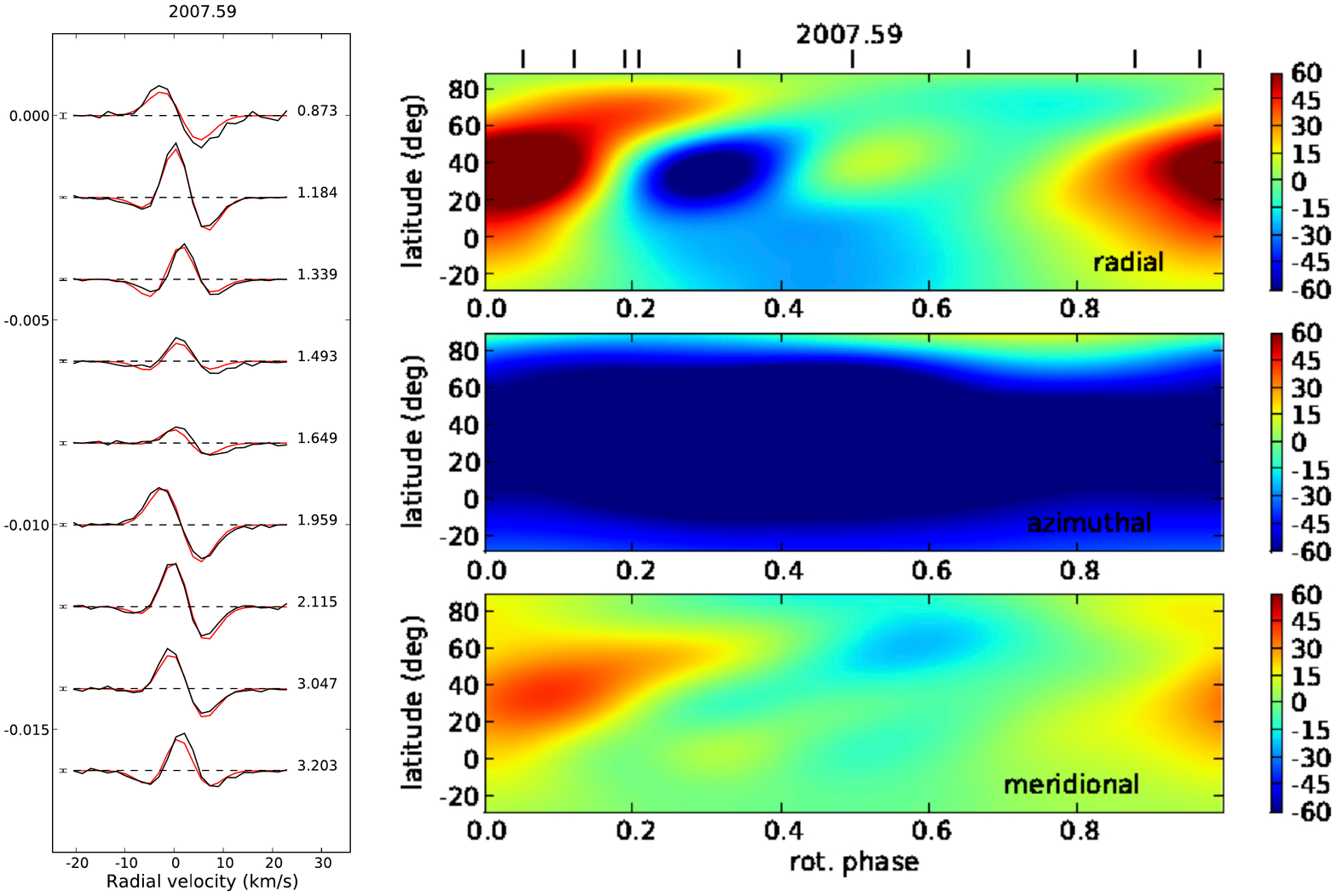}
\caption[]{Circular polarisation profiles (left panel) and magnetic map (right panel) of $\xi$~Boo~A in July 2007 from \cite{2012A&A...540A.138M}.}
\label{fig:ksiboo}
\end{figure}

Sun-like stars do not all exhibit the same magnetic field characteristics. For example, the G dwarf $\xi$~Bootis~A - $\rm T_{eff} = 5570$~K, $\rm M\sim 0.86 M_{\odot}$, $\rm v sini = 3.0 \pm 0.5$~km s$^{-1}$ (\cite[Valenti \& Fischer 2005]{2005yCat..21590141V}), $\rm P_{rot} = 6.4$~days (\cite[Toner \& Gray 1988]{1988ApJ...334.1008T}) - observed in July 2007 (\cite[Morgenthaler et al. 2011, 2012]{2011AN....332..866M,2012A&A...540A.138M}), shows a stronger magnetic field (80 G) with 80\% of the energy in the toroidal component of the field. The circular polarisation profiles and the reconstructed magnetic map are shown in Fig. \ref{fig:ksiboo}.

\begin{figure}
\centering
	\includegraphics[scale=0.4]{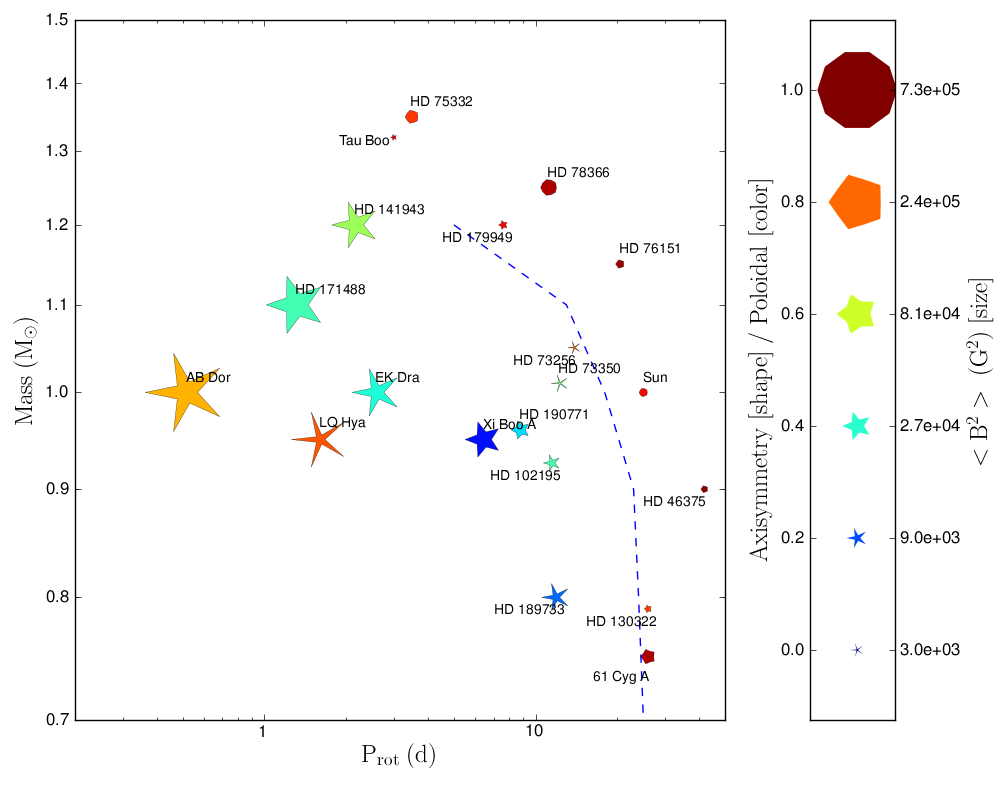}
\caption[]{Mass-rotation diagram of reconstructed magnetic fields of Sun-like stars. The dashed line shows Rossby number of unity. The size of the symbol represents the field strength, its color the contribution of the poloidal component to the field, and its shape the degree axisymmetry of the poloidal component.}
\label{fig:confuso}
\end{figure}

Those two previous examples show that there is a variety of magnetic field topologies and strengths among Sun-like stars. One can examine if there is a trend with stellar properties (e.g. rotation, temperature, mass). To answer this question, we have adapted the graph of Fig. 3 in \cite{2009ARA&A..47..333D} to include all the data for Sun-like stars. Fig. \ref{fig:confuso} represents the general characteristics of the reconstructed magnetic fields: the reconstructed magnetic energy density (i.e. the integral of $\langle B^2 \rangle$ over the surface), the percentage of poloidal field, and finally the fractional energy density in axisymmetric modes (i.e., with $m\textless l/2$, m and l being the order and degree of the spherical harmonic modes describing the reconstructed field). Each observed star for which a magnetic map has been reconstructed is indicated by a symbol at its rotation period and mass, with the symbol size reflecting the magnetic energy density, the symbol color reflecting if the field is mainly poloidal or mainly toroidal, and the symbol shape indicating how axisymmetric the poloidal component is. The data are from \cite[Catala et al.(2007); Moutou et al. (2007); Donati et al. (2003, 2008a); Petit et al. (2008, 2009); Jeffers \& Donati(2008); Donati \& Landstreet(2009); Fares et al.(2009,2010,2012,2013); Morgenthaler et al.(2011,2012) and Marsden et al.(2011) ]{2009ARA&A..47..333D,2007MNRAS.374L..42C,2008MNRAS.385.1179D,2003MNRAS.345.1145D,
2013MNRAS.tmp.2010F,2012MNRAS.423.1006F,
2010MNRAS.406..409F,2009MNRAS.398.1383F,2007A&A...473..651M,
2008MNRAS.388...80P,2008MNRAS.390..635J,2011MNRAS.413.1922M,2011AN....332..866M,2012A&A...540A.138M,2009A&A...508L...9P}.

Fig. \ref{fig:confuso} shows a variety of observed topologies. The fields of these stars can be either poloidal or toroidal, axisymmetric or non-axisymmetric, and of different strengths. The field strengths are in general smaller than those of M dwarfs (see \cite[Reiners \& Basri 2006, 2010; Morin et al. 2008, 2010 and Donati et al. 2008b]{2006ApJ...644..497R,2010ApJ...710..924R,2008MNRAS.390..567M,2010MNRAS.407.2269M,
2008MNRAS.390..545D})

However, if we overplot the Rossby number equal to one, we see the main trend (the Rossby number is the ratio of the rotation period of the star to the convective turnover time):

stars having a Rossby number greater than one seem to have a weak, mainly poloidal and axisymmetric magnetic fields; while stars with Rossby number smaller than one (in the range of masses considered in this work) seem to have a stronger, mainly toroidal and non-axisymmetric magnetic fields.

\section{Magnetic Cycles}

The Sun goes through activity and magnetic cycles, as discussed in Section \ref{sec:intro}. Monitoring the activity and searching for cycles of Sun-like stars has been the subject of many studies (e.g. \cite[Henry et al 1996; Baliunas et al. 1995, 1997]{1996AJ....111..439H,1995ApJ...438..269B,1997ApJ...474L.119B}). Some stars exhibit activity cycles, which vary in duration between almost a year to 25 years (e.g. \cite[Metcalfe et al. 2010]{2010ApJ...723L.213M}). In addition, asteroseismology can also reveal activity cycles, by measuring the variation of the amplitude of the modes and the frequency shifts (e.g. \cite[Garc{\'{\i}}a et al. 2010]{2010Sci...329.1032G}). 

\begin{figure*}
\includegraphics[scale=0.7]{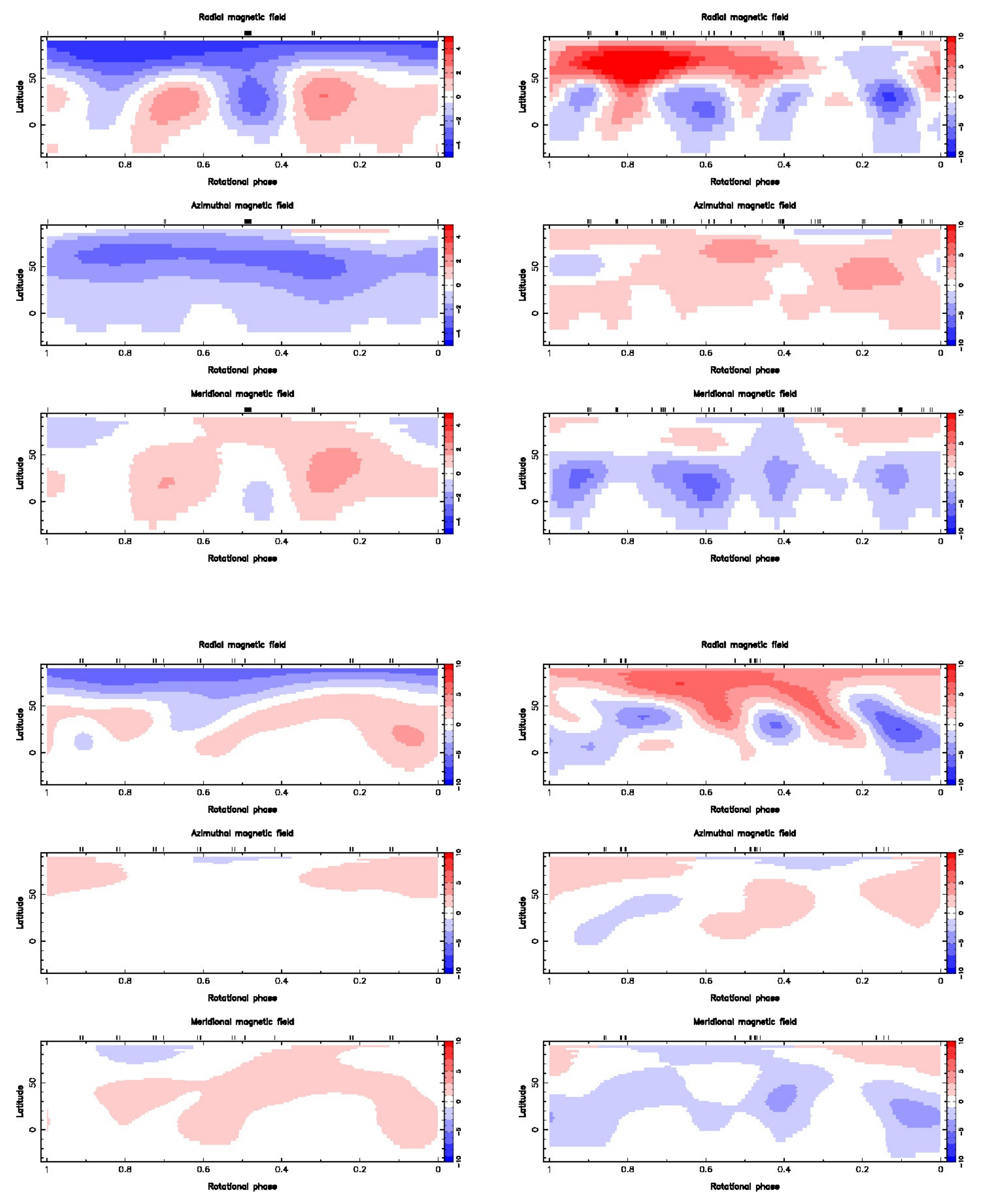}
\caption{Magnetic maps of $\tau$~Boo. Top left: June 2006; top right: June 2007; bottom left: May 2008; bottom right July 2009. The maps show polarity flips every year, observed in all components of the field. Data adapted from \cite[Catala et al. (2007); Donati et al. (2008a); Fares et al. (2009,2013)]{2007MNRAS.374L..42C,2008MNRAS.385.1179D,2009MNRAS.398.1383F,2013MNRAS.tmp.2010F}.}
\label{fig:tauboo}
\end{figure*}

Knowing that activity cycles exist for Sun-like stars, one can ask if large-scale magnetic cycles exist as well. 
In order to investigate that, some stars were monitored over many epochs of observation. For example, $\tau$~Boo, an F star with $\rm T_{eff} = 6387$~K, $\rm M\sim1.33M_{\odot}$, $\rm  v sini = 15.0 \pm 0.5$~km s$^{-1}$ (\cite[Valenti \& Fischer 2005]{2005yCat..21590141V}),
$\rm P_{rot} = 3.0 - 3.9$~days, $\rm d\Omega = 0.4$~rad day$^{-1}$ (\cite[Fares et al. 2009]{2009MNRAS.398.1383F}), and host to a hot-Jupiter,  has shown a reversal of the polarity of the polar field every year between 2006 and 2009 (see Fig. \ref{fig:tauboo}). The star exhibits strong differential rotation, is orbited by a very massive planet, and has a rotation period similar to the planetary orbital period (\cite[Fares et al. 2009]{2009MNRAS.398.1383F}). The magnetic field is mainly poloidal and axisymmetric for these epochs, with a mean field strength of about 3-5 G.

The star was also observed in January 2008 half way between the reversal. The field is mainly azimuthal for this epoch (toroidal field). The field thus changes from a mainly poloidal to a mainly toroidal configuration, and than goes back to the poloidal configuration but with a different polarity. The magnetic cycle for this star is of 2 years, however a shorter cycle of 8 months cannot be ruled out.
The effect of tidal interactions between the massive planet and the shallow outer convective envelope of the star was suggested as a possible cause for the short magnetic cycle. However, more recent studies have found a fast magnetic cycle of 3 years on another F star HD78366 (\cite[Morgenthaler et al. 2011]{2011AN....332..866M}). 

\subsection*{Complex Cycle}

\begin{figure*}
\includegraphics[height=\textwidth]{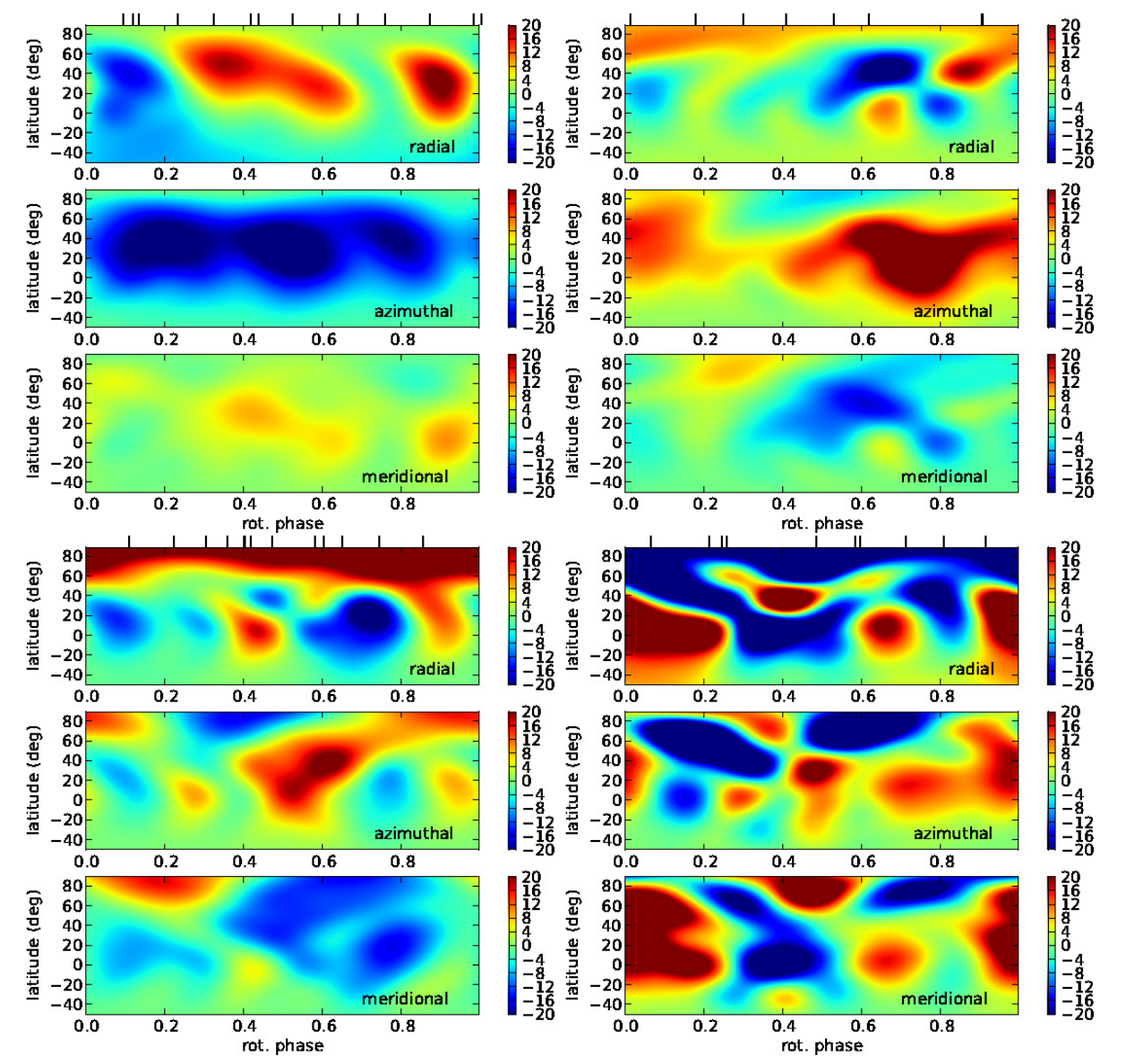}
\caption{Magnetic maps of HD~190771 for 2007.59, 2008.67, 2009.47, and 2010.50 data sets
from left to right and top to bottom (\cite[Morgenthaler et al. 2011]{2011AN....332..866M}).}
\label{fig:hd190771}
\end{figure*}

The magnetic field evolution of stars does not always show a simple change in polarity. For instance, HD190771 shows a complex cycle ($\rm T_{eff} = 5834$~K, $\rm M \sim 0.96 M_{\odot}$, $\rm v sini = 4.3 \pm 0.5$~km/s (\cite[Valenti \& Fischer 2005]{2005yCat..21590141V}), $\rm P_{rot} = 8.8$~days (\cite[Toner \& Gray 1988]{1988ApJ...334.1008T}), $\rm d\Omega = 0.12$~rad/d (\cite[Petit et al. 2009]{2009A&A...508L...9P}). This star was observed every year from 2007 until 2011 (\cite{2009A&A...508L...9P,2011AN....332..866M} and Petit private com). The azimuthal field changes polarity between 2007 and 2008 (see Fig. \ref{fig:hd190771}), the field has a simple topology for both epochs. However, the topology of the field becomes more complex after 2008. A polarity switch is observed in 2010 when compared to 2008, but the field is much more complex in 2010. In 2011, the field goes back to a simple topology, having the same polarity as it had in 2008. The magnetic cycle of this star is complex when compared to the solar cycle.

The magnetic fields of Sun-like stars evolve with time. Stars that do not exhibit magnetic cycles still show an evolution of the poloidal and toroidal components of the field. Yet they do not show a polarity switch, e.g. $\xi$~Boo~A (\cite[Morgenthaler et al. 2012]{2012A&A...540A.138M}). These observations, in case of failure in detecting magnetic polarity flips, indicate that `failed cycles' can also happen, i.e. variability without polarity flips. 

\subsection*{Stellar Cycles versus the Solar Cycle}

For the Sun, the length of the activity cycle is half that of the magnetic cycle. One can question if this relation holds for stars, and what we can learn about the generation of their magnetic fields.

Let us consider a stellar magnetic cycle with polarity flips, like the one of $\tau$~Boo. Although it is a fast cycle when compared to the solar cycle, it is considered normal since it shows regular polarity flips. This star was observed by the HK project of Mount Wilson. It has a long activity cycle of ~11.6 years (\cite[Baliunas et al. 1995]{1995ApJ...438..269B}). If we compare this to the solar case, the activity cycle is much longer than the magnetic cycle. This system is an interesting one. \cite{2000ApJ...531..415H} found a persistent 116 day period over 30 years of observations in the CaII fluxes. However, this period does not appear in radial velocity nor photometric data. They say that it cannot be explained by the familiar phenomena of rotation, growth and decay of surface features, or an activity cycle. However, the spectropolarimetric data cannot rule out a magnetic cycle of 240 days (8 months; \cite[Fares et al. 2013]{2013MNRAS.tmp.2010F}).  If this is the real period of the cycle, not only is it much shorter than the solar one, but it is also almost twice the 116 day period found in the CaII data. 

\cite{2012AN....333...26P} attempted to detect an X-ray cycle for $\tau$~Boo by observing the star over 6 epochs. Although the star exhibits variability in X-ray, it does not show a cyclic behavior. However, the lack of an X-ray cycle is not inconsistent with the existence of a mangetic cycle. Theoretical work shows that a magnetic cycle does not necessarily imply an X-ray one (e.g. \cite[McIvor et al. 2006]{2006MNRAS.367..592M}). Also, \cite[Vidotto et al.(2012)]{2012MNRAS.423.3285V} simulated its stellar wind through the magnetic cycle, using the reconstructed maps as boundary conditions for the stellar magnetic field. They calculated the X-ray emission measure and find that this does not vary during the cycle, which agrees with the findings of \cite{2012AN....333...26P}.

The relations between activity cycles, magnetic cycles, and X-ray cycles should thus be investigated for stars. The picture we have currently is not similar to the solar one. Understanding the difference will improve our understanding on magnetic field generation. 

\section{Conclusions}

The study of the magnetic fields of Sun-like stars gives new insights into stellar magnetism and provides constraints for dynamo theories. The magnetic fields in these stars are generated in the outer convective layers. However, a full understanding of the dynamo mechanisms acting in these layers has not yet been reached.

Sun-like stars have a wide range of magnetic properties. Magnetic field strengths and geometries vary between mainly poloidal and mainly toroidal fields. Field strengths vary between a few Gauss to a few hundred Gauss. 
Despite this variety,  there are trends with stellar properties, especially the Rossby number. Sun-like stars with Rossby numbers smaller than unity have mainly toroidal magnetic fields, while stars with Rossby numbers greater than unity seem to have weaker magnetic fields, dominated by the poloidal components.

However, one should consider the existence of magnetic cycle. Some Sun-like stars do indeed have cycles; we observe solar-like cycles with polarity flips, more complex cycles, and some cyclic variations without polarity flips. The length of the current discovered magnetic cycles was surprising, as these cycles are very short when compared to the solar cycle.

This leads us to a set of important questions: is the solar cycle unusual? How do magnetic and activity cycles correlate in Sun-like stars? What are the stellar characteristics that drive the cycles (differential rotation, rotation,...)? Observations and theoretical works are still needed to answer these open questions.

\section*{Acknowledgments}
RF acknowledges support from the STFC.

\end{document}